# FAPS: A Fast Platform for Protein Structureomics Analysis

*Lucas Wilken,[1] Nihjum Paul,[2] Troy Timmerman,[3] Sara A. Tolba,[3] Amara Arshad,[4] Di Wu,[1] Wenjie Xia,[6] Bakhtiyor Rasulev,[3,5] Rick Jansen,[7] and Dali Sun [8,9]*\**

[1] Department of Electrical and Computer Engineering, North Dakota State University, 1411 Centennial Blvd., 101S Fargo, ND 58102.
[2] Department of Public Health, Genomics, Phenomics, and Bioinformatics Program, North Dakota State University, 1455 14th Ave, Fargo, ND 58102.
[3] Materials and Nanotechnology Program, North Dakota State University, 1410 North 14th Avenue, CIE 201, Fargo, ND 58102.
[4] Department of Civil, Construction and Environmental Engineering, North Dakota State University, 1410 North 14th Avenue, CIE 201, Fargo, ND 58102.
[5] Department of Coatings and Polymeric Materials, College of Science and Mathematics, North Dakota State University, Fargo, ND 58102.
[6] Department of Aerospace Engineering, Iowa State University, Ames, IA 50011, USA.
[7] Clinical and Translational Science Institute, 717 Delaware Street SE, Second Floor, Minneapolis, MN 55414
[8] Biomedical Focus, Department of Electrical and Computer Engineering, University of Denver, 2155 E Wesley Ave., Denver, CO 80208, USA
[9] Knoebel Institute of Healthy Aging, University of Denver, 2155 E Wesley Ave., Denver, CO 80208, USA

*Corresponding author: Dali Sun, E-mail address: dali.sun@du.edu

**Abstract**

Protein quantification and analysis are well-accepted approaches for biomarker discovery but are limited to identification without structural information. High-throughput omics data (i.e., genomics, transcriptomics, and proteomics) have become pervasive in cancer biology studies and reach well beyond more specialized areas such as metabolomics, epigenomics, pharmacogenomics, and interact-omics. However, large-scale analysis based on the structure of the biomolecules, namely structure-omics, is still underexplored due to a lack of handy tools. In response, we developed the Fast Analysis of Protein Structure (FAPS) database, a platform designed to advance quantitative proteomics to structure-omics analysis, which significantly shortens large-scale structure-omics from weeks to seconds. FAPS can serve as a new protein secondary structure database, providing a centralized and functional database for both simulated and experimentally determined bioinformatics statistics relating to secondary structure. Stored data is generated both through the structure simulation, currently SWISS-MODEL and AlphaFold, performed by high-performance computers, and the pre-existing UniProt database. FAPS provides user-friendly features that create a straightforward and effective way of accessing accurate data on the proportion of secondary structure in different protein chains, providing a fast numerical and visual reference for protein structure calculations and analysis. FAPS is accessible through http://fapsdb.org.

## 1. Background

Modern mass spectrometry-based protein profiling introduces not only individual identification, but also quantitative information of individual proteins contained in a sample. This contributes to comparative proteomics to enumerate the proteins in malignant and nonmalignant tissue cells. Protein quantification and analysis are well-accepted approaches for biomarker discovery but limited to identification without structural information [1]. Biomarkers are increasingly used to improve patient diagnosis and monitor therapeutic responses [2]. A number of omics methods, such as genomics, transcriptomics, proteomics, and metabolomics, hold special promise for the discovery of novel cancer biomarkers that might form the foundation for new clinical blood tests [3], but a long-standing problem with current cancer marker studies is that seldom of these markers have been deemed adequate for clinical applications in detection, measuring patient response to drugs, understanding tumor heterogeneity, or discriminating against different diseases [4]. This reveals the deficiency of the traditional marker discovery pipeline based on identification and quantification without structural information.

Most biomarker studies focus on a single molecular marker (protein/RNA/DNA based) or signatures based on multiplex assays [5]. However, their effectiveness is dissatisfactory for clinical prognosis or diagnosis, which motivated us to study them in an alternative way that considers collective attributes rather than as a single marker [6]. The secondary structure of the tumorous proteins is a collective attribute different from the conventional single marker discovery process. Superior to the primary sequence, tertiary and quaternary structure of proteins, which require intricate measurement apparatus, the methods to evaluate the secondary structure are comparatively more accessible for clinical translation. Several experimental methods are currently used to determine the structure of a protein, including X-ray diffraction (XRD), nuclear magnetic resonance (NMR), circular dichroism (CD), Fourier-transform infrared spectroscopy (FTIR), Raman, and electron microscopy [7–9]. Each method has advantages and disadvantages, but the substantial time consumed for experimental structural identification presents significant translational barriers for large-scale structure-omics analysis. That is why none of the well-known protein databases (e.g., Entrez [10], UniProt [11], Swissprot [12]) have embraced full storage of structural information for the entire human proteome. A convenient tool enabling fast structural analysis will break the technical barrier for collective stereochemical biomarker exploration and facilitate biological and stereopharmacology development for cancer.

In order to introduce structural information into marker discovery, intact protein structural information must be obtained. However, none of the current protein structural databases fulfill this request. The Entrez databases, as an example, host 874,272 protein entries, but only 167,650 (<20%) possess experimentally determined structural information, and not all these records present intact protein structure. Thus, structural prediction based on simulation would be necessary to cover the deficiencies of these protein structure databases. However, integrating database searches with the known structural and simulated structure suffers from a timing mismatch. The time cost of simulation algorithms for predicting individual proteins is at a level of minutes. However, one structural database search is in a time scale of seconds. None of the currently available platforms solve this time mismatching [13].

Previously, we studied the collective attributes [14] of tumor cells by analyzing the secondary structure of cellular proteins, and found malignant cells are richer in β-sheet structure than their nonmalignant counterparts [6]. We also designed a *de novo* workflow combining proteomics, bioinformatics and protein folding simulations to study collective attributes, counting individual proteins' contributions. This novel workflow greatly extended the proteomics study to collective structural analysis, enabling the structural signature of the tumor cells. However, the flow involved a time-consuming simulation module that predicted the structures of the proteins without experimentally determined structures. To adapt large-scale structure-omics analysis, we initiated the Fast Analysis of Protein Structure (FAPS) database to store structural information from the experimentally determined structure database (UniProt) and simulation. FAPS enabled three dimensions (identification, quantification, and structural information) for marker discovery instead of the current 2D (identification, quantification) methodology. The new tool integrates a database containing experimentally determined structural information from UniProt, and pre-simulated the structure of unknown partial of the proteins, which solves the mismatching and empowers fast analysis at a level of seconds. A biologist can easily access the web interface (http://fapsdb.org/) to obtain structural details quickly.

FAPS is designed as a protein bioinformatics tool to simplify the retrieval, visualization, and analysis of protein structure. Data generated both through computational simulation and through experimental bioinformatics analysis is stored and processed in a straightforward format to streamline user accessibility. These data structure formats are stored both in raw formats and in generated visualizations, to serve both analytical and fast reference needs.

## 2. Construction and Utilities

FAPS was designed to fill both data generation and data access needs; to that end, it was necessary to build appropriate structures for both purposes. Throughout the design process of the FAPS system, user experience and access efficiency were kept at the forefront, with the goal of providing a quick retrieval service to the end user.

### 2.1 System Design

The database storage/generation system is built to be used and operated on a two-sided model of design, as shown in Fig.1. The prediction modules (SWISS-MODEL and AlphaFold) check the corresponding database entry (downloaded from https://swissmodel.expasy.org/repository, and https://alphafold.ebi.ac.uk/download) for existing records; if not found, it will run the corresponding simulation program on the high-performance computing clusters (HPC) [15] to predict the structure model. Existing structural information from UniProt and data generated by simulation are glued, and the overall constitute is statistically calculated by the HPC. Briefly, proteins are searched for in the protein databank (UniProt) and the secondary structure information is extracted. The primary sequence from the database is used in structure prediction simulation to estimate the structural percentage. Since most of the proteins in the protein bank do not have completed secondary structure information, the final structural percentage ($SP$) of the protein was estimated by the following equation, $SP = pSP_{db} + (1-p)SP_{sim}$, where $p$ is the percentage of the sequence length with a known structure in the protein bank, $SP_{db}$ and $SP_{sim}$ are the structural features' (secondary) percentages calculated from the protein database and simulation, respectively. All scripts on HPC are written in Python. The calculated proportion is then sent to the second pillar of the system, the web server and database, as JSON data over the internet and processed for storage and display.

The calculated structure feature information is packed into JSON data and sent to Heroku, a platform as a service (PaaS) that enables developers to build, run, and operate applications entirely in the cloud [16]. The data is then stored in the PostgreSQL database hosted at Heroku, where the web server is also hosted. Django, an open-source web framework [17], is utilized to design and construct the web application. The user is then able to query the database to find the structure compositions of specific protein accession ID, or through an uploaded list.

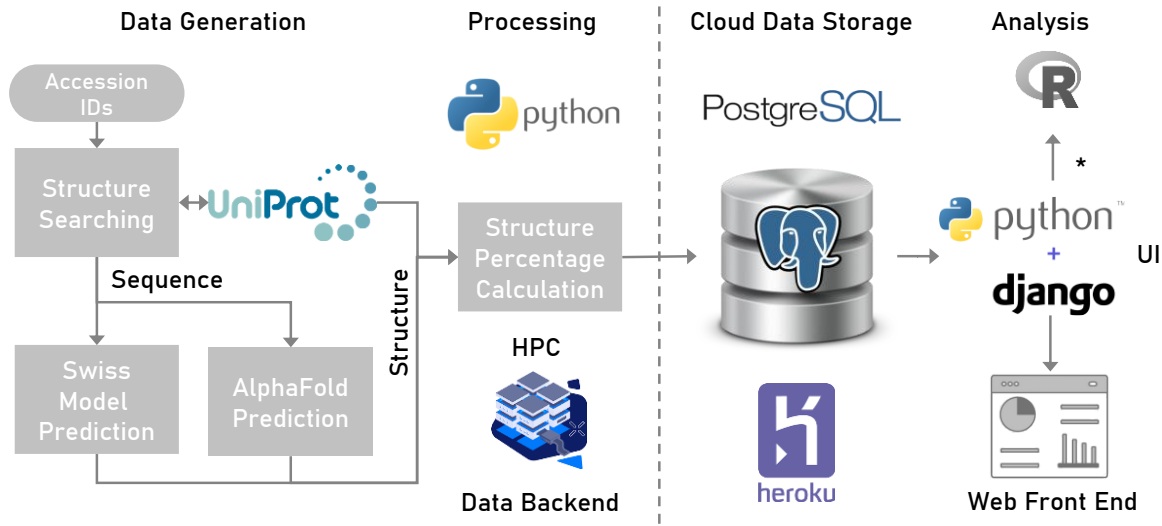

Figure 1. FAPS DB System Flow. * prospective functions.

## 2.2 Database Design

The primary concern of FAPS's design is speed and simplicity. Since the data to store is well structured and must be highly scalable to ensure fast query results, a relational database was selected to store the data. PostgreSQL is a powerful, open-source object-relational database system with high reliability, feature robustness, and performance [18]. As such, PostgreSQL was selected for fast queries, reduced footprint, and strong data integrity semantics.

The database scheme is shown in Fig. 2. Structured JSON data from HPC consisting of a simulation or UniProt protein model is stored in 'simProtein' or 'uniProtein' table, respectively. Secondary structure proportions, accession code, sequence length, and the unknown and known proportions of the structure are stored in each table depending on the data type. These tables are foreign keyed to a 'masterProtein' model, which is used to optimize and simplify queries as a linking table. When a user executes a search, only the 'masterProtein' needs to be interrogated to return both results, halving the search time. Simulation data is linked in a many-to-one relationship to the master, as there could potentially be multiple simulation types for each protein. UniProt data is linked in a one-to-one relationship to avoid duplicate information being stored. Currently, the database stores approximately 1500 entries.

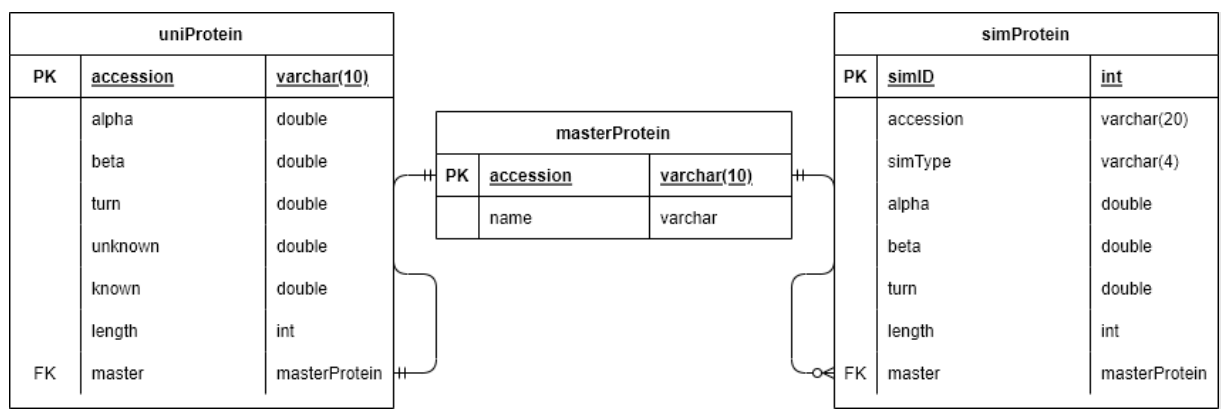

Figure 2. Database Entity Relationship Diagram (ERD)

## 2.3 User Interface Design

To enhance clarity, simplicity, and consistency, Bootstrap, a web user interface library developed by Meta, was implemented at the front-end of the user interface [19]. The app back-end was framed by Django [17], a web framework for Python. For better data visualization, several popular open-sources interactive visualization libraries will be integrated into the front-end (i.e., D3.js [20], Google Charts [21]). Following the biologists-centered design principle, the web interface design is simple and informative with pages shown in Fig. 3A-C, and the user flow also is designed for a human-friendly intuitive path (Fig.3D). To guide first-time users, a step-by-step help page is available for reference should there be issues with queries. User feedback is also considered – the website supports a user suggestion and feedback submission form, where features can be requested, or bugs reported. The tools/modules used in the development are listed in Table 1.

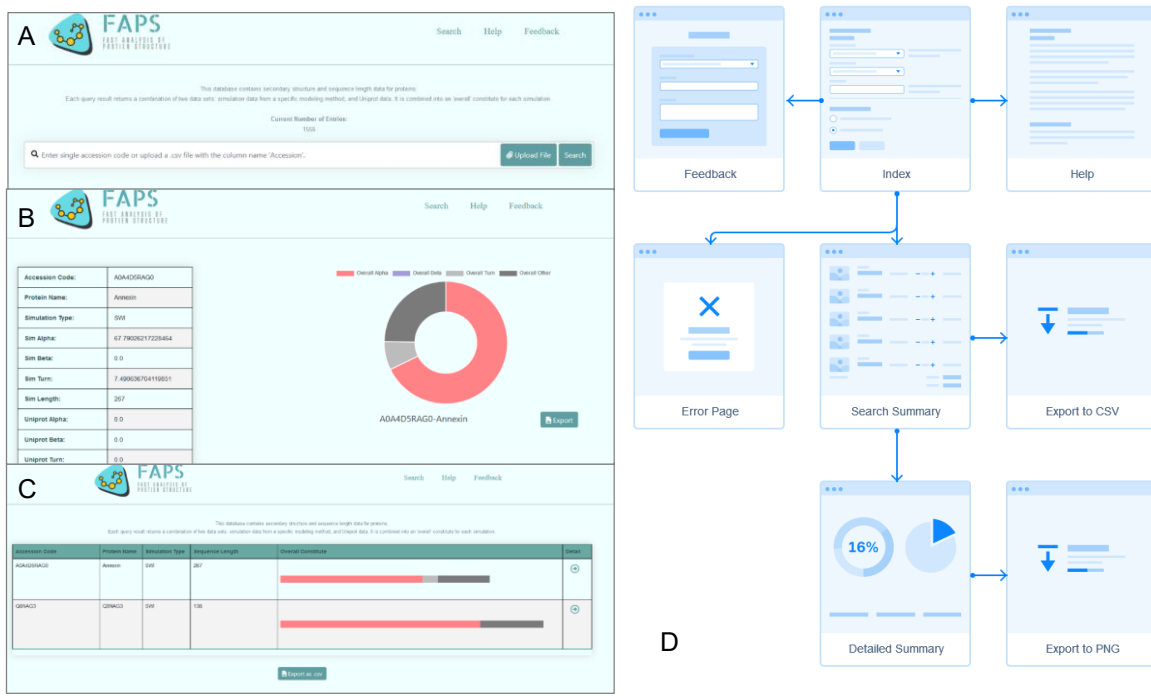

Figure 3. FAPS Web Interface (A,B,C) and user flow design (D)

As shown in Fig. 4, users can access and retrieve data through the FAPS webpage by searching for their specific accession code or through a list from an .csv file. For instance, if the user was to enter the partial accession code 'AG' or the full code 'A0A4D5RAG0,' they would be prompted with the table in Fig. 3B first and then the expanded detail view in Fig. 3B. All searches return these two key results: a summarizing data table of raw proportional and length data for each matched accession code, and a more in-depth detail page with a protein visualization pie chart. These can be exported to a .csv file and a .png image, respectively, so the website user can store and reference them.

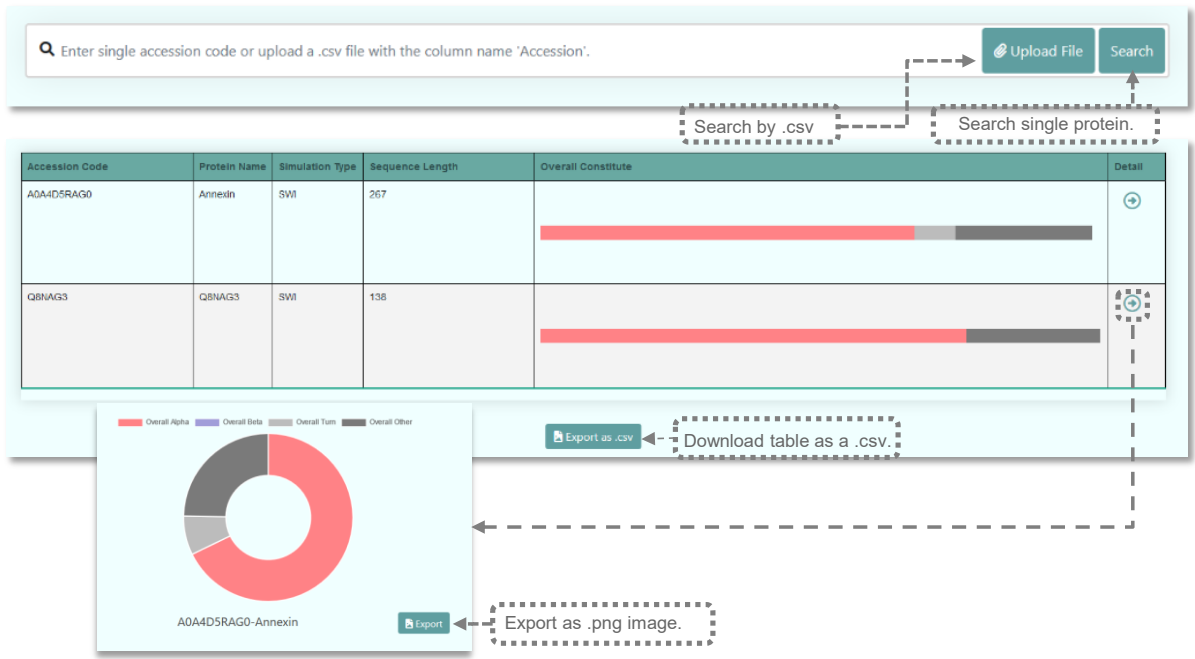

Figure 4. A typical use case

Overall, the tools and resources embedded within FAPS create an efficient and effective user experience and analytical process. Ease-of-use is substantially emphasized, and high-intensity calculations can be saved to reduce redundant searching.

*Table 1: Tools/modules integrated in development*

| NAME | DESCRIPTION |
|---|---|
| ALPHAFOLD | DeepMind that predicts a protein's 3D structure |
| SWISS MODEL | A protein structure homology-modeling program |
| BEAUTIFUL SOUP 4 | A Python library for pulling data out of HTML and XML files |
| BOOTSTRAP | Web user interface library developed by Facebook. |
| BIOPYTHON | Freely available tools for biological computation |
| D3.JS | JavaScript library for Data Visualization |
| DJANGO | Web framework for Python |
| GOOGLE CHART | Interactive library for Data Visualization |
| POSTGRESQL | Relational Database |
| REACT TABLE | DataGrid library for interactive table |

## 2.4 UniProt Integration

For scraping data from UniProt on HPC, the Python package 'Beautiful Soup 4' was used in tandem with the 'requests' package; these are used to send HTTP requests to the https://www.uniprot.org/uniprot/ website [22]. The secondary structure table is extracted from the UniProt webpage, which contains the length of the alpha helix, beta sheet, and random coil structure parameters (Fig. 5).

From this table, six parameters are calculated: helix percentage (hp), known partial helix percentage (kphp), beta-strand percentage (bp), known partial beta strand percentage (kpbp), random coil percentage (rp), and known partial random coil percentage (kprp). Each of the structure's proportion is calculated by dividing its total length by the sum of all parameter's length. The known partial of each parameter is calculated by dividing their total length by the length of known structure for that accession code. The script outputs a .csv file for each accession code containing its corresponding parameters, containing seven columns.

As this scraping script was also designed to work on HPC, all errors that occurred while running the program are logged into a text file for analysis. The 'logging' package was used for this purpose. The log file contains the error type, error name, and the date and time the error occurred.

The .csv file generated from the web scraping is then fed to FAPS database under the 'uniProtein' table through a pre-prepared upload package. This upload package converts the generated .csv file into web JSON data, which is sent to the FAPS website back-end with the aforementioned 'requests' package once the proper credentials have been uploaded and interpreted. The JSON data is then interpreted and converted into a 'uniProtein' entry in the database.

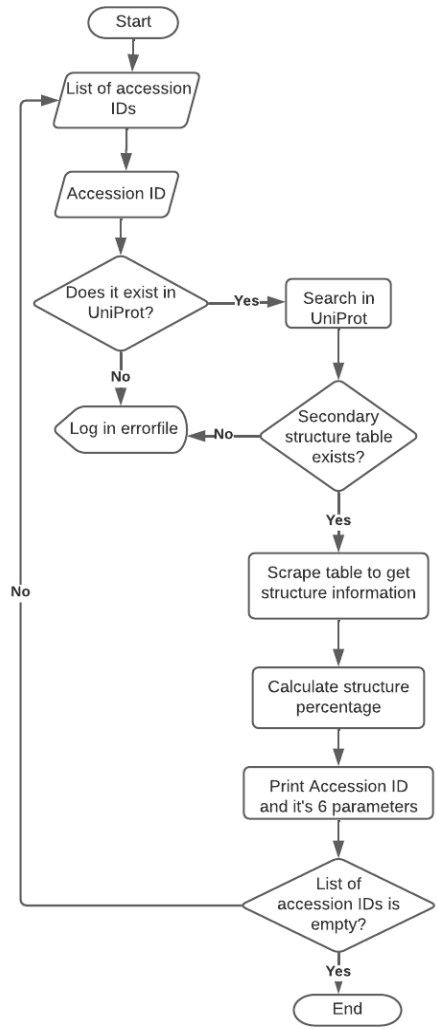

Figure 5. Flow of UniProt Integration

## 2.5 Simulation Integration: SWISS-MODEL

Since most of the proteins do not have determined structures, FAPS integrates simulated structure information in calculations to estimate protein secondary structure where experimental structural data from UniProt might not exist. To utilize simulations, a program was designed on HPC which integrates the SWISS-MODEL simulation with the FAPS upload scheme. SWISS-MODEL is a leading homology for the simulation and modeling of 3D protein structure, available online for use from a remote server in which model results are emailed to the user [23]. This is problematic when simulations need to be done in large batch sizes, which necessitates the creation of a localized server for the use of the SWISS-MODEL system.

Key considerations were given to the batch size and time constraints. SWISS-MODEL takes an average of 2-10 minutes to generate the protein model when run on HPC. UniProt accession codes are given as input, and a target model with secondary structure information is given as output. Because the goal of FAPS is to handle massive protein quantities, input is accepted from a list (.csv) file, in which UniProt accession codes are stored in a column with the header 'Accession'.

The program takes the .csv input and looks for a column containing UniProt accession codes[22]. The primary sequence of the individual proteins is then retrieved from UniProt's repository. This sequence is then used to query the SWISS-MODEL database for homologous protein models. After finding matches, it generates a multiple sequence alignment with the top template candidate, which is then used to create the model. The model is optimized using the Promod3 model engine [24]. The secondary structure proportions are then calculated using the Dictionary of Protein Secondary Structure (DSS) [25]. The process is repeated for each submitted protein. Finally, the simulation data is pushed to the FAPS database ('simProtein' table) with 'SWI' tag using a modified version of the upload script specified in the previous section.

## 2.6    Simulation Integration: AlphaFold

With the advent of artificial intelligence (AI) algorithms, it becomes possible to assume that a structure, either computationally generated or experimentally deciphered, is at researchers' fingertips for nearly every protein. AlphaFold also hits the secondary structure prediction to identify novel targets with limited or no structure information [26]. Compared to the huge database repositories of the popular homology modeling program SWISS-MODEL, AlphaFold has established huge database repositories of modeled 3D structures and supplemented them with prediction algorithms for the annotation of secondary structures, protein domains and functions. Deep machine learning algorithms were also introduced to the 3D structure prediction, to increase the accuracy of models significantly [27].

AlphaFold incorporates the novice neural network architectures with training procedures that exploit biological, physical, and evolutionary knowledge into geometric constraints of 3D protein structure, leveraging sequence into a deep learning algorithm. The AlphaFold network predicts the precise 3D coordinates of all atoms by using aligned amino acid sequences of given homologous proteins as inputs [28].

The AlphaFold prediction is also used in secondary structure calculations where experimental data from UniProt and homology data from SWISS-MODEL might not be available to predict reliable 3D structures. AlphaFold has greatly changed the structure prediction research method from low throughput to a high throughput manner. We utilize the high throughput power of AlphaFold with HPC resources to explore the potential of optimizing time constraints for large-scale proteins with the FAPS database arrangements.

UniProt accession codes are used to fetch the amino acid sequence in the FASTA format from Uniprot's database utilized by the AlphaFold; as a result, a target model is achieved, with high confidence. The query sequence is passed through the novice repeated layers of a neural network as multiple sequence alignments and takes templates from AlphaFold database to predict the optimized 3D structure. At the end, the predicted structures are passed through a DSSP (hydrogen bond estimation) algorithm to assign the secondary structure at each amino acid level. We implement the high throughput end of AlphaFold with GPUs to predict large datasets of proteins within a few minutes. Final secondary structure data is exported based on the above sections to the FAPS database ('simProtein' table) with 'AF' simulation type tag.

## 3. Discussion and Conclusion

To promote the adoption of large-scale structural analysis, we designed the FAPS database to store structural information from UniProt and simulations. A biologist can easily access the web interface (http://fapsdb.org/) to promptly obtain structural details. FAPS combines experiment-determined and simulated structure information, and none of the current bioinformatics databases or tools provide alternative solutions for intact large-scale structural information that FAPS is capable of. The pre-calculated database innovatively reduces large-scale structure-omics analysis from weeks/months/years to seconds. FAPS, together with quantitative proteomics, enables three-dimension analysis (identification, quantification, and structural information) for marker discovery instead of the current two-dimension (identification, quantification) methodology. It also highlighted an underexplored biomarker category: stereochemical biomarkers. The tool makes the collective stereochemical biomarker discovery feasible, which potentiates further cancer study (i.e., stereopharmacology, stereoselectivity). The FAPS database project provides users with user-friendly reference tools to streamline the process of protein secondary structure analysis. This has applications in proteomics, bioinformatics, and cancer research, where the secondary structure of a particular species of protein is paramount in analyzing cell and amino acid interactions.

In future efforts, we will further develop innovative informatics methods and tools which facilitate cancer biology, studying and highlighting collective stereochemical signatures as new biomarkers. We will develop the tools (e.g., *R* interface) while keeping the focus on the simplicity of the user interface and user flow, to ensure easy access from the end-user biologist to the tool without complex training. FAPS may be expanded to support different simulation and modeling pipelines; the long-term goal is to promote the ability of structure-omics to impact cancer research and the development of relevant therapeutic strategies. The results derived from the project will be able to provide structural signatures of tumorous proteins, which highlights a new approach to stereopharmacology. The project will substantially reduce barriers for biomedical researchers in mining complex functional proteomic data, serve as a hub for large-scale structure-omics, and directly facilitate new cancer biology and marker discovery.

The FAPS is available for access at http://fapsdb.org.

**List of Abbreviations**

FAPS (DB) – Fast Analysis of Protein Structure (Database)

XRD – X-ray Diffraction

NMR – Nuclear Magnetic Resonance Spectroscopy

CD – Circular Dichroism

FTIR – Fourier-Transform Infrared Spectroscopy

HPC – High Performance Computing

SP – Structure Percentage

JSON – JavaScript Object Notation

HTTP – Hypertext Transfer Protocol

DSS – Dictionary of Secondary Structure

GPU – Graphics Processing Unit

**Declarations**

**Competing interests**

Authors declare that they have no competing interests.

**Funding**

This work was financially supported by grants from the National Cancer Institute (R03CA252783, R21CA270748) and the National Institute of General Medical Sciences (U54GM128729) of National Institutes of Health to DS, NDSU EPSCoR STEM Research and Education fund (FAR0032086) to DS, ND EPSCoR: Advancing Science Excellence in ND (FAR0030554) to DS, National Science Foundation (NSF) under NSF EPSCoR Track-1 Cooperative Agreement (OIA #1355466) to DS. NSF under NSF OIA ND-ACES (Award #1946202) to WX, NDSU Foundation and Alumni Association to DS. This work was partly supported by the National Science Foundation NSF-MRI award OAC-2019077.

**Authors' contributions**

DS conceived the idea, provided financial support, and supervised the project. LW, AA, NP, TT, and DS wrote the manuscript. LW, NP, and SS constructed the website. DS, RJ, BR, WX, and DW contributed to editing the manuscript. TT, BR, AA and WX contributed to molecular simulation. DS approved the final version of the manuscript.

**Acknowledgments**

The authors acknowledge North Dakota State University Center for Computationally Assisted Science and Technology for computing resources.